\documentclass[technote]{IEEEtran}
\usepackage{cite,units}
\ifCLASSINFOpdf
   \usepackage[pdftex]{graphicx}
  \DeclareGraphicsExtensions{.pdf,.jpeg,.png,.eps}
\else
  \usepackage[dvips]{graphicx}
   \graphicspath{{../eps/}}
  \DeclareGraphicsExtensions{.eps}
\fi
\usepackage{array}
\usepackage[tight,footnotesize]{subfigure}
\usepackage[cmex10]{amsmath}
\usepackage{cleveref}
\usepackage{mdwmath}
\usepackage{mdwtab}

\crefname{equation}{equation}{equations}
\Crefname{equation}{Equation}{Equations}
\crefrangelabelformat{equation}{(#3#1#4--#5#2#6)}
\crefmultiformat{equation}{equations (#2#1#3}{, #2#1#3)}{#2#1#3}{#2#1#3}
\Crefmultiformat{equation}{Equations (#2#1#3}{, #2#1#3)}{#2#1#3}{#2#1#3}
\allowdisplaybreaks[3]
\begin{document}
\title{Miniaturized Circular-Waveguide Probe Antennas Using Metamaterial Liners}
\author{Justin~G.~Pollock,~\IEEEmembership{Graduate~Student~Member,~IEEE,}
        and~Ashwin~K.~Iyer,~\IEEEmembership{Senior Member,~IEEE}%
}
\markboth{Submitted to the IEEE Transactions on Antennas and Propagation}%
{Shell \MakeLowercase{\textit{et al.}}: IEEE Journals}
\maketitle
\begin{abstract}
This work presents the radiation performance of open-ended circular-waveguide probe antennas that have been miniaturized by the introduction of thin metamaterial liners. The liners introduce an {\em HE}$_{11}$ mode well below the natural cutoff frequency, which provides substantial gain improvements over a similarly sized waveguide probe. A new feeding arrangement employing a shielded-loop source embedded inside the miniaturized waveguide is developed to efficiently excite the {\em HE}$_{11}$ mode and avoid the excitation of other modes across the frequency-reduced band while maintaining the antenna's compactness. A metamaterial-lined circular-waveguide probe antenna operating over $42\%$ below its natural cutoff frequency is designed to provide a radiation efficiency of up to $28.8\%$. A simple, printed-circuit implementation of the metamaterial liner based on inductively loaded wires is proposed and its dispersion features are discussed.\end{abstract}

\section{Introduction}
\label{sec:intro}

\IEEEPARstart{O}PEN-ended waveguide (OEWG) probe antennas are popular for their well known near- and far-field patterns, ease of integration with standardized waveguides, and established calibration techniques. In the microwave and millimeter-wave regimes, waveguides are recognized for their applications in non-destructive dielectric sensing~\cite{decreton1975nondestructive}, surface crack detection~\cite{huber1997modeling}, thermography~\cite{barrett1986basic}, and near-field antenna and material characterization~\cite{yaghjian1986overview}. In the latter, an OEWG probe is used to sample spatial electromagnetic field profiles, in which the measured signal is related to an analog integration of the fields across its aperture. However, these probes only operate efficiently for frequencies above a cutoff frequency, $f_c$, which can lead to a few inherent drawbacks~\cite{tice1955probes}. For instance, conventional above-cutoff waveguides are large and unable to sample the fields to even moderately sub-wavelength resolutions; moreover, they can unnecessarily load the environment, potentially introducing errors into the measurements. Decreasing the aperture size of an OEWG probe would present several benefits, provided its efficiency could be maintained. One of these is the mitigation of errors introduced by approximations in near-field probe corrections~\cite{hansen1988spherical}.

Whereas OEWG probes may be miniaturized by operating in their evanescent region (i.e., below their fundamental cutoff frequency), such evanescent waveguide antennas require additional design, such as the inclusion of dielectric regions or complex matching networks, to compensate for their naturally reactive guided-wave impedance. For instance, the evanescent rectangular waveguide antenna reported in Ref.~\cite{ludlow2013applying} requires multiple capacitive posts to achieve a $-10$dB bandwidth of $15\%$ with a gain less than $4.7$dB. Furthermore, this band resides immediately below its fundamental cutoff, implying only moderate or no miniaturization. Operating more deeply into cutoff to achieve larger degrees of miniaturization would result in severe decreases in gain and bandwidth. Another approach is to homogeneously fill the waveguide with a high-permittivity dielectric to reduce its cutoff frequency; however, at low frequencies, the dielectric filling leads to further increases in weight and manufacturing cost. Furthermore, this approach is unsuitable for applications requiring access to the interior of the waveguide. For example, uninhibited motion of particles through waveguides is required in both microwave heating of fluids~\cite{ratanadecho2002numerical} and in cyclotron masers for electron beams~\cite{yin1993cyclotron}. In these cases, it may be more prudent to partially fill, or line, the waveguide with a dielectric~\cite{kildal2005}.
\begin{figure}[!t]
\centering
\includegraphics[width=1.6in]{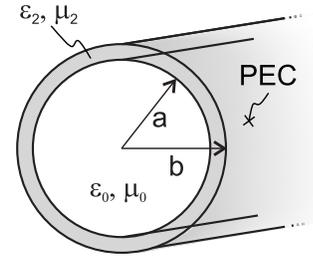}
\caption{A PEC circular waveguide consisting of two concentric dielectric regions: an inner vacuum region of radius $a$ and an outer dielectric region of thickness $t=b-a$ described by permittivity and permeability $\epsilon_{2}$ and $\mu_{2}$.\label{fig0}}
\end{figure}

Consider the PEC circular waveguide shown in Fig.~\ref{fig0} that contains two concentric regions: an inner vacuum region and an outer dielectric region, or liner. Whereas thin liners composed of conventional positive-permittivity dielectrics may not allow for any substantial reduction in cutoff frequency, it was recently shown that the inclusion of thin \textit{metamaterial} liners into PEC circular waveguides can permit propagation well below the unlined waveguide's fundamental-mode cutoff~\cite{pollockmttt2013}. In that work, a frequency-reduced backward-wave band corresponding to a hybrid electric {\em HE}$_{11}$ mode occurred in the regime in which the metamaterial liner exhibited dispersive epsilon-negative and near-zero (ENNZ) properties, over $42\%$ below its fundamental-mode cutoff frequency. This mode exhibits uniform, strongly collimated fields, with most of the field variation taking place in the liner. Reference~\cite{pollockmttt2013} also developed a methodology to design for an arbitrarily reduced cutoff frequency, where the most dramatic reduction is achieved for extreme ENNZ values (i.e., as $\epsilon\rightarrow 0^-$).

This work explores the application of metamaterial-lined circular waveguides to the design of miniaturized OEWG probe antennas~\cite{pollockapsursi2014}. Section~\ref{sec:theory} presents the dispersion properties of a suite of frequency-reduced modes supported by these miniaturized waveguide sections. Section~\ref{sec:sims} compares the radiation characteristics of the {\em HE}$_{11}$ and {\em TE}$_{11}$ modes of miniaturized and conventional-sized OEWGs, respectively, using full-wave simulations. A novel shielded-loop source is developed to practically excite the {\em HE}$_{11}$ mode in the metamaterial-lined OEWG while suppressing other modes, and the resulting patterns and radiation efficiency are investigated. In Sec.~\ref{sec:realization}, a practical implementation of ENNZ liners based on easily-fabricated printed-circuit layers is proposed and is shown to exhibit a frequency-reduced {\em HE}$_{11}$ passband.
\section{Theory}
Figure~\ref{fig0} presents the geometry of the open-ended metamaterial-lined circular waveguide (radius $b$) under consideration. An inner-core vacuum region (radius $a$) of permittivity $\epsilon_0$ and permeability $\mu_0$ is surrounded by a metamaterial layer of thickness $t=b-a$ with a non-magnetic permeability ($\mu_{2}=\mu_0$) and dispersive permittivity $\epsilon_{2}(f)$. The complex permittivity dispersion is described by a Drude model with $\epsilon_{2}(f)=\epsilon_0[1-f_{ep}^2/f(f-jf_t)]$, in which $f_{ep}$ is the plasma frequency and $f_t$ is the damping frequency establishing the liner's loss. The waveguide dimensions are chosen to be $b=15$mm and $a=14$mm with liner Drude-model parameters $f_{ep}=3.550$GHz and $f_t=0$MHz (i.e., lossless).
\label{sec:theory}
\begin{figure}[!t]
\centering
\includegraphics[width=3in]{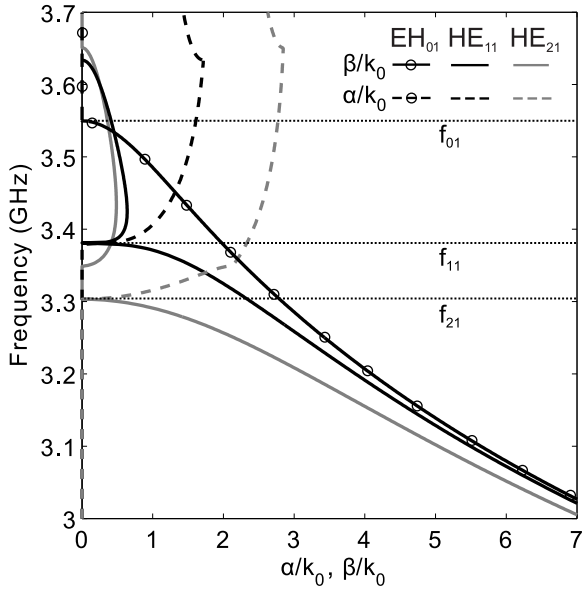}
\caption{\unskip Dispersion of $\alpha/k_0$ (dashed curves) and $\beta/k_0$ (solid curves) for the isotropic metamaterial-lined waveguide's {\em HE}$_{11}$ (black curves), {\em HE}$_{21}$ (grey curves), and {\em EH}$_{01}$ (black circle markers) modes.\label{fig1:subfig2}}
\vspace{-5pt}
\end{figure}

The solution of the dispersion equation~\cite{pollockmttt2013} reveals several modes below the natural {\em TE}$_{11}$ cutoff ($5.861$GHz). Figure~\ref{fig1:subfig2} focuses on the frequency-reduced {\em EH}$_{01}$ (black circle markers), {\em HE}$_{11}$ (black curves), and {\em HE}$_{21}$ (grey curves) modes assuming no losses. The solid and dashed lines, respectively, indicate $\beta/k_0$ and $\alpha/k_0$. The cutoff frequencies of the frequency-reduced {\em EH}$_{01}$, {\em HE}$_{11}$ and {\em HE}$_{21}$ modes' propagating bands are $f_{01}=3.550$GHz, $f_{11}=3.381$GHz, and $f_{21}=3.303$GHz, respectively. These lie in the ENNZ region of $\epsilon_{2}(f)$ and are $42\%$--$66\%$ below their corresponding natural cutoffs. These modes exhibit a backward-wave trend, for which large phase shifts are achieved at lower frequencies. As such, the {\em EH}$_{01}$ propagating band ($f<f_{01}$) overlaps that of the other modes. Additionally, each of the {\em HE}$_{11}$ and {\em HE}$_{21}$ modes exhibit two regions of interest: a propagating backward-wave band ($f<f_{11}$ and $f<f_{21}$, respectively) and also a complex-propagation band that does not allow for transmission of real power~\cite{chorney1961power}. The inclusion of low losses, such as those observed in transmission-line metamaterials~\cite{pollockmttt2013}, does not significantly alter the dispersion of $\beta$, but does introduce a non-negligible $\alpha$. However, $\alpha$ remains much smaller than $\beta$ well into the backward-wave band.
\section{Full-Wave Simulations}
\label{sec:sims}
\subsection{Waveguide Excitation}
\begin{figure}[!t]
\centering
\subfigure[]{
\includegraphics[width=2.8in]{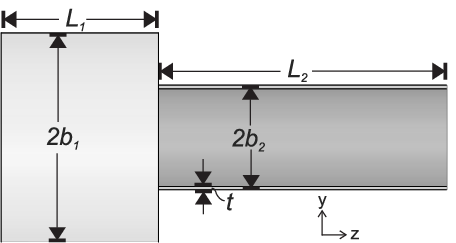}
\label{fig3:subfig1}
}
\subfigure[]{
\includegraphics[width=3in]{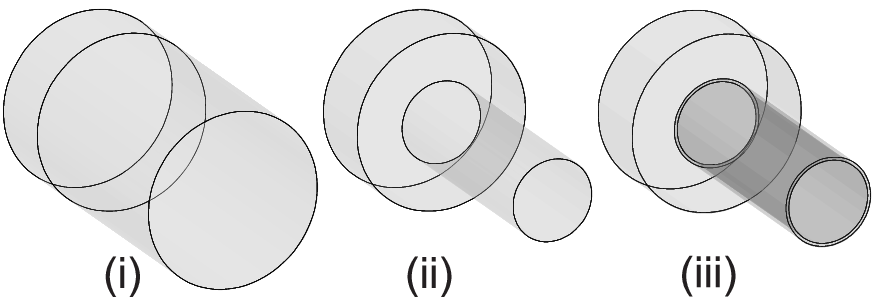}
\label{fig3:subfig2}
}
\subfigure[]{
\includegraphics[width=3.2in]{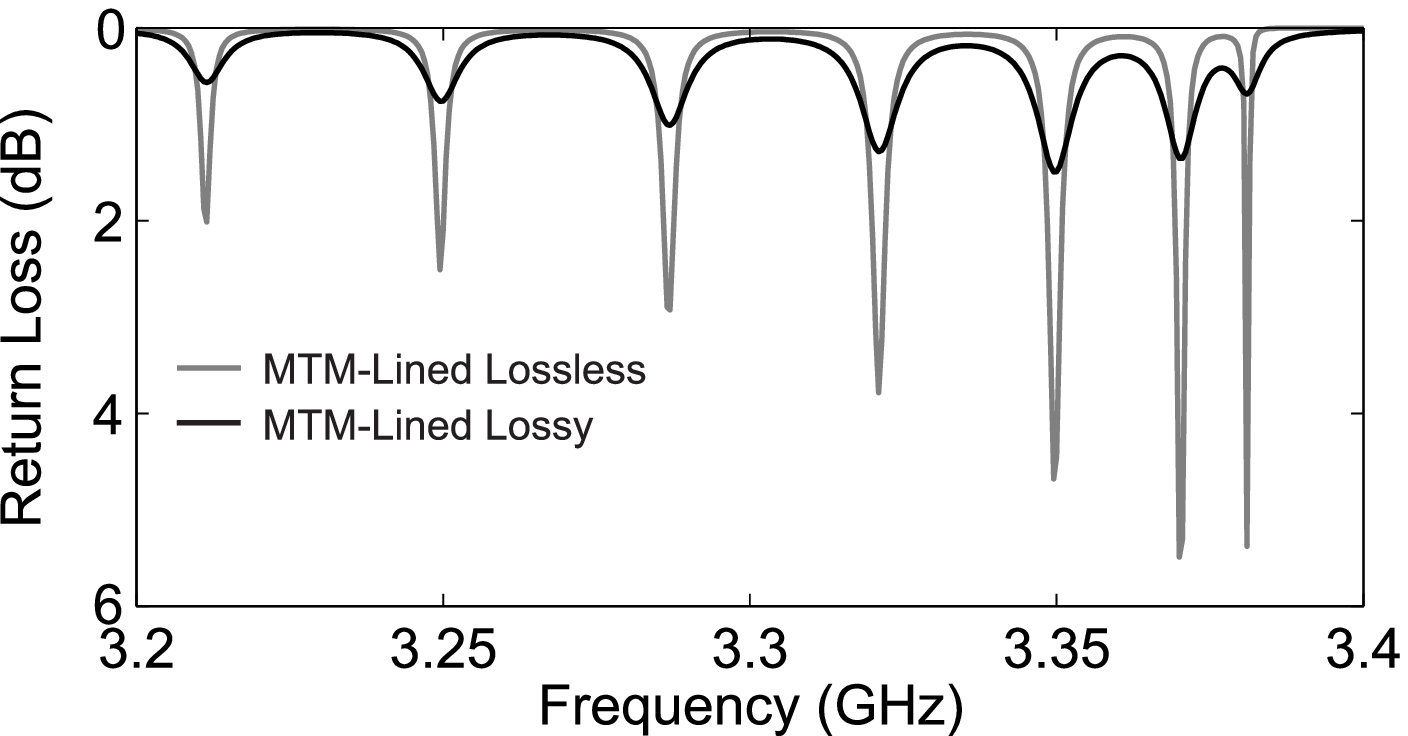}
\label{fig3:subfig3}
}
\caption{\unskip(a) Generic full-wave simulation model of the OEWG probe used in the comparison of the three cases shown in (b). Case (i) is a large above-cutoff waveguide and termed `Standard'. Case (ii) is an above-cutoff waveguide connected to a smaller below-cutoff waveguide and termed `Cutoff'. In case (iii), the smaller waveguide contains a metamaterial liner possessing complex dispersive permittivity and is termed `MTM-Lined'. (c) The return loss for the `MTM-Lined Lossless' (grey curve) and `MTM-Lined Lossy' (black curve) OEWGs.}
\vspace{-5pt}
\end{figure}

In the previous section, it was shown that metamaterial liners introduce a below-cutoff {\em HE}$_{11}$-mode backward-wave passband in the liner's ENNZ range. Furthermore, Ref.~\cite{pollockmttt2013} established that this band is capable of transmitting power at levels well above those of a similarly sized unlined (and below-cutoff) waveguide. In this section, it will be demonstrated that such miniaturized waveguides are capable of efficiently radiating power into free space, and their directivity patterns are similar to those of the homogeneously vacuum-filled OEWG. The full-wave simulation software HFSS~\cite{hfss} is employed to simulate the generic model shown in Fig.~\ref{fig3:subfig1}, for the three cases shown in Fig.~\ref{fig3:subfig2}. In each case, a large vacuum-filled PEC circular waveguide of length $L_1=20$mm and radius $b_1=30$mm contains a waveport at its closed end that excites the {\em TE}$_{11}$ mode for frequencies above its cutoff of $2.930$GHz. This, in turn, couples into a connecting waveguide section of length $L_2=82.5$mm that is opened at one end, and the whole structure is enclosed in a spherical radiation boundary of radius $76$mm.  The use of a discontinuous waveguide junction is based on similar setups employed in Refs.~\cite{pollockmttt2013,enghati_tunnel} to observe transmission through below-cutoff waveguides and to limit the study to particular modes of interest.
\begin{figure}[!t]
\centering
\subfigure[]{
\includegraphics[width=1.5in]{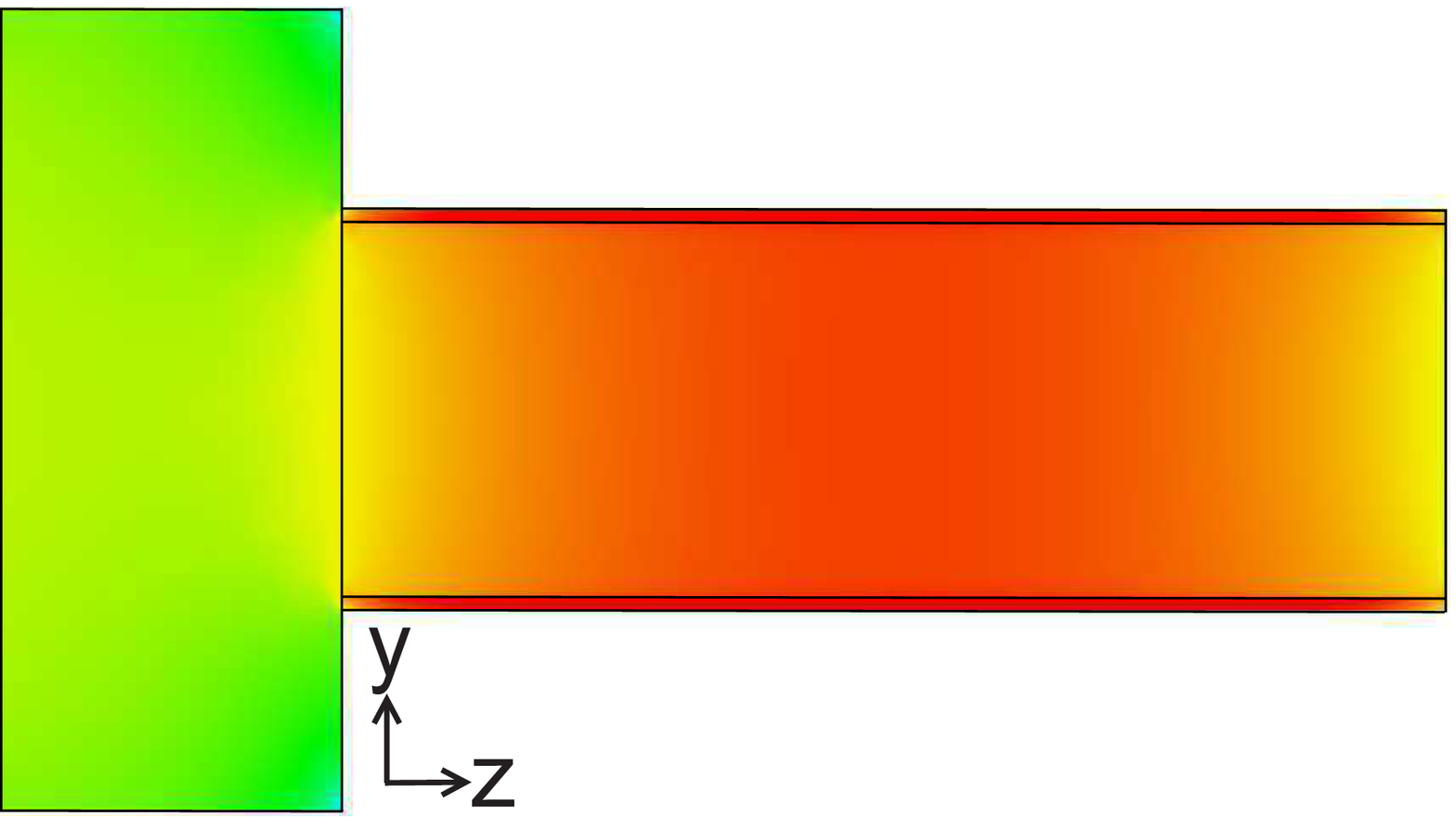}
\label{fig4:subfig1}
}
\subfigure[]{
\includegraphics[width=1.5in]{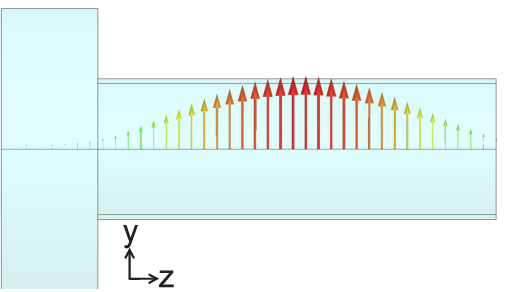}
\label{fig4:subfig2}
}
\subfigure[]{
\includegraphics[width=1.35in]{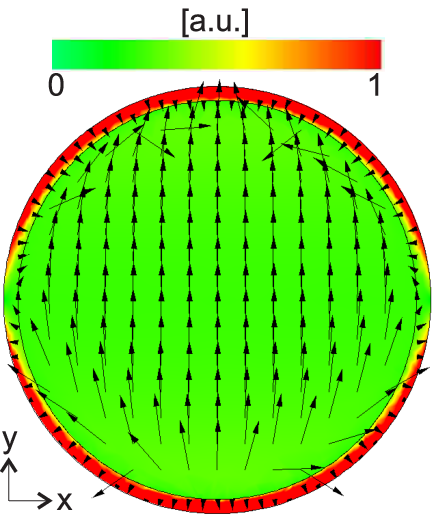}
\label{fig4:subfig3}
}
\subfigure[]{
\includegraphics[width=1.35in]{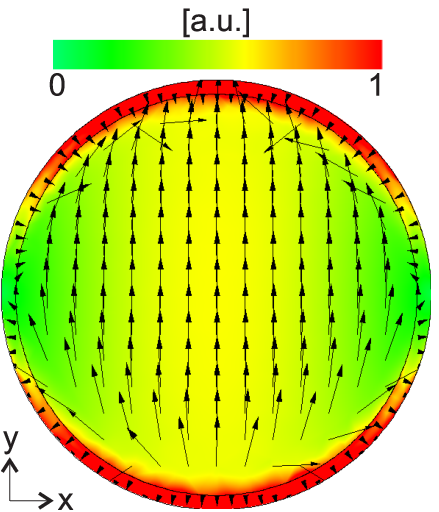}
\label{fig4:subfig4}
}
\caption{\unskip\unskip Complex electric field (a) magnitudes and (b) vectors along a longitudinal cut in the E-plane. Electric field vectors along a transverse cut taken at (c) the center of the `MTM-lined' OEWG and (d) its open aperture. All data presented at $f=3.381$GHz for the lossless case.}
\vspace{-5pt}
\end{figure}
Now, we establish nomenclature to distinguish the different connecting OEWGs in each case in Fig.~\ref{fig3:subfig2}. Case (i) contains a homogeneously vacuum-filled OEWG of radius equal to that of the excitation waveguide ($b_2=30$mm) operating above its {\em TE}$_{11}$ mode's natural cutoff frequency of $f_c=2.930$GHz and is termed `Standard'. Case (ii) contains a homogeneously vacuum-filled OEWG of half the radius ($b_2=15$mm) operating below its natural cutoff frequency of $f_c=5.861$GHz and is termed `Cutoff'. Case (iii) lines the under-cutoff waveguide of case (ii) with an ENNZ metamaterial, whose dispersion and geometrical parameters ($b_2=15$mm, $t=1$mm) were presented in Sec.~\ref{sec:theory}, and is termed `MTM-Lined'.

The metamaterial liner is assigned the dispersive permittivity, $\epsilon_{2}(f)$, reported in Sec.~\ref{sec:theory} for both lossless (termed `MTM-Lined Lossless' with $f_t=0$MHz) and lossy (termed `MTM-Lined Lossy' with $f_t=5$MHz) cases. According to Fig.~\ref{fig1:subfig2}, {\em HE}$_{11}$ propagation should occur below $f_{11}=3.381$GHz for the frequency-reduced backward-wave band. Figure~\ref{fig3:subfig3} shows the return loss in the `MTM-Lined Lossless' (grey curve) and `MTM-Lined Lossy' (black curve) cases. Indeed, the metamaterial liner results in a passband, which is described by several resonant return-loss peaks. As previously shown in Ref.~\cite{pollockmttt2013}, each of these peaks is akin to a Fabry-P\'{e}rot-type resonance through the metamaterial-lined waveguide section. The distribution of resonance frequencies may be designed by varying either the metamaterial-lined waveguide's length or the liner's permittivity dispersion profile. Figures~\ref{fig4:subfig1}-\ref{fig4:subfig2} present the complex electric-field magnitudes and vectors, respectively, in the E-plane ($\phi=90^{\circ}$) at $f=3.381$GHz, in which the resonant distribution is evident. The electric fields are purely transverse ($\rho$- and $\phi$-directed) with no longitudinal ($z$-directed) component, which verifies that an {\em HE}-type mode, and not the {\em EH}$_{01}$ mode, is strongly excited at the waveguide junction.  An examination of the fields in the transverse plane at the center of the `MTM-lined' OEWG (Fig.~\ref{fig4:subfig3}) and at its open aperture (Fig.~\ref{fig4:subfig4}) reveals a distinct {\em HE}$_{11}$ field pattern, as desired.
\begin{table}[t]
\begin{center}\label{table:1}
\begin{tabular}{lcccc}
\multicolumn{5}{c}{TABLE I: \textsc{Antenna Parameters}} \\
 &Standard& Cutoff&\multicolumn{2}{c}{MTM-Lined}\\
 &&& Lossless &Lossy\\
\hline
 Freq. (GHz)&3.381&3.381&3.381&3.350\\
 Direct. (dB)&7.69&5.96&5.49&5.32\\
 $\eta_{rad}$ ($\%$)&96&0&75.3&12.9\\
 Gain (dB)&7.51&-71.5&4.26&-3.57\\
 HPBW$_{H}$($^\circ$)&69.0&82.0&77.0&76.4\\
 HPBW$_{E}$($^\circ$)&71.0&74.8&72.6&72.2\\
 F/B (dB) &8.54&5.33&3.83&3.86\\
\lasthline
\end{tabular}
\end{center}
\end{table}
\begin{figure}[!t]
\centering
\subfigure[]{
\includegraphics[width=1.7in]{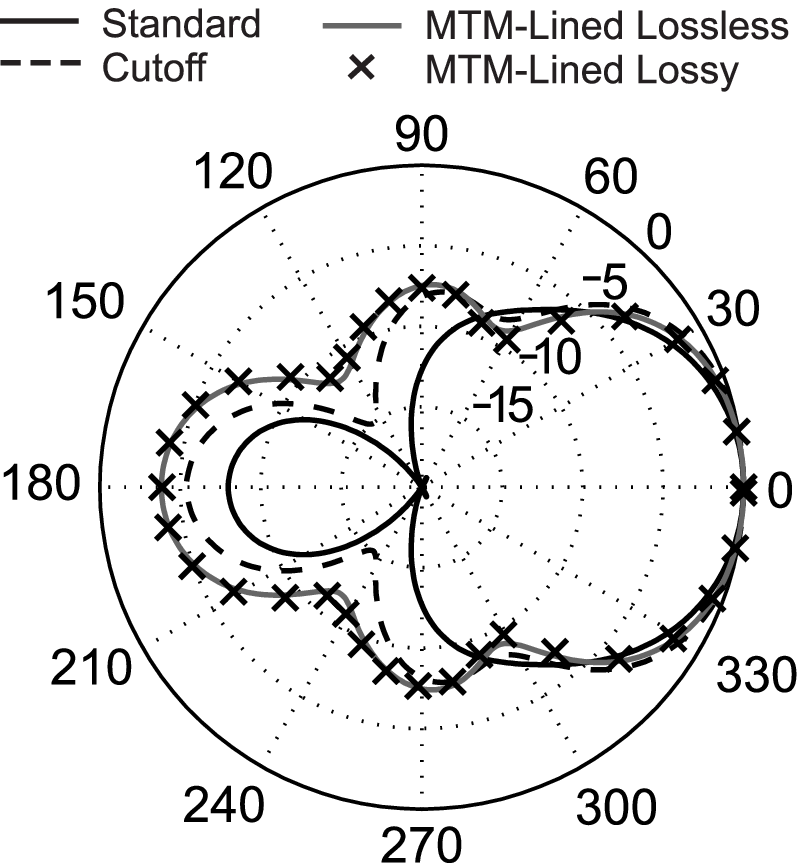}
\label{fig5:subfig1}
}
\subfigure[]{
\includegraphics[width=1.7in]{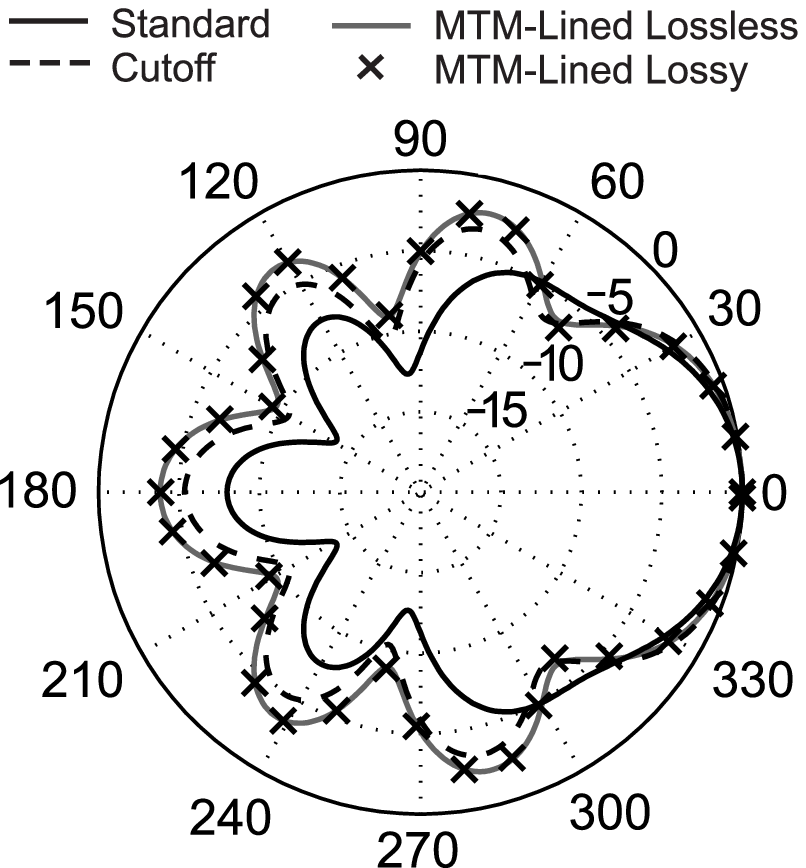}
\label{fig5:subfig2}
}
\caption{\unskip\unskip Comparing normalized (a) H-plane and (b) E-plane directivity patterns of the OEWG obtained from simulations of the `Standard', `Cutoff', and `MTM-Lined' (Lossless and Lossy) cases. All data presented at $f=3.381$GHz.}
\vspace{-5pt}
\end{figure}

Figures~\ref{fig5:subfig1}-\ref{fig5:subfig2} present the H-plane ($\phi=0^{\circ}$) and E-plane ($\phi=90^{\circ}$) cuts, respectively, of the normalized directivity patterns for each investigated case. In addition, Table I presents relevant antenna parameters such as radiation efficiency ($\eta_{rad}$), half-power beamwidth (HPBW), and front-to-back (F/B) ratio. At $f=3.381$GHz, the `Standard' (solid black curve) case achieves a $7.69$dB directivity, an H-plane HPBW of $69.0^{\circ}$, and a respectable F/B ratio of $8.54$dB. Decreasing the outer radius by half in the `Cutoff' (dashed black curve) case causes a $1.73$dB decrease in directivity and a $13.0^{\circ}$ increase in the H-plane HPBW. An increase in the side-lobe and back-lobe levels suggests that substantial currents are also formed on the outer surface of the PEC waveguides. However, as indicated in Table I, the operation of the `Cutoff' waveguide in the evanescent regime results in truly negligible radiation efficiency and hence, no useful gain.

Now, introducing the lossless metamaterial liner into the `Cutoff' case and focusing on the radiation pattern at its first resonant peak at $f=3.381$GHz (solid grey curve), the directivity is roughly maintained at $5.49$dB. However, the improved return loss at this frequency due to the metamaterial liner causes a dramatic increase in the radiation efficiency ($75.3\%$) and gain ($4.26$dB). Since the `MTM-Lined Lossless' and `Lossy' liners have similar aperture-field profiles, the pattern in the lossy case (black markers) overlaps that of the lossless case. However, the introduction of loss reduces the relative aperture-field strengths, and for the third resonant peak at $f=3.350$GHz, the radiation efficiency is only $12.9\%$ corresponding to a gain of $-3.57$dB. This is, nevertheless, a useful gain that represents a nearly $70$dB improvement over the `Cutoff' OEWG. Such gain improvements may potentially enable previously impossible sub-wavelength spatial resolution above the noise floor, and/or with increased dynamic range, particularly in continuous-wave applications or those involving narrowband phenomena.

Due in part to reflections at the open aperture that couple back into the waveport, this manner of excitation suffers from poor matching. The next section investigates a practical source to more efficiently excite the {\em HE}$_{11}$ mode and to mitigate the potential excitation of other modes. Additionally, this design eliminates the need for a larger excitation waveguide and enhances the compactness of the OEWG probe.
\begin{figure}[!t]
\centering
\subfigure[]{
\includegraphics[width=2.5in]{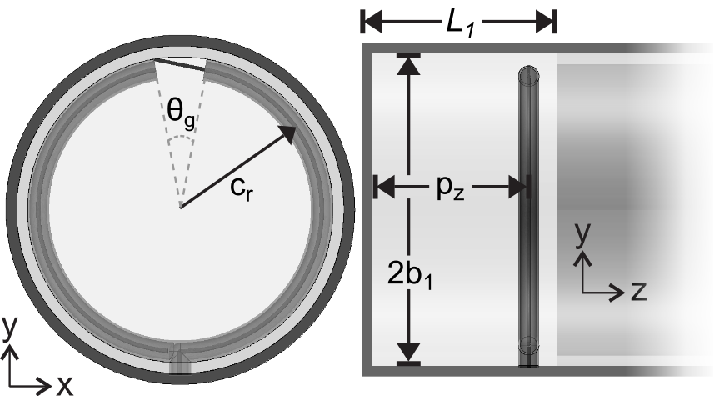}
\label{fig6:subfig1}
}
\\
\subfigure[]{
\includegraphics[width=3.2in]{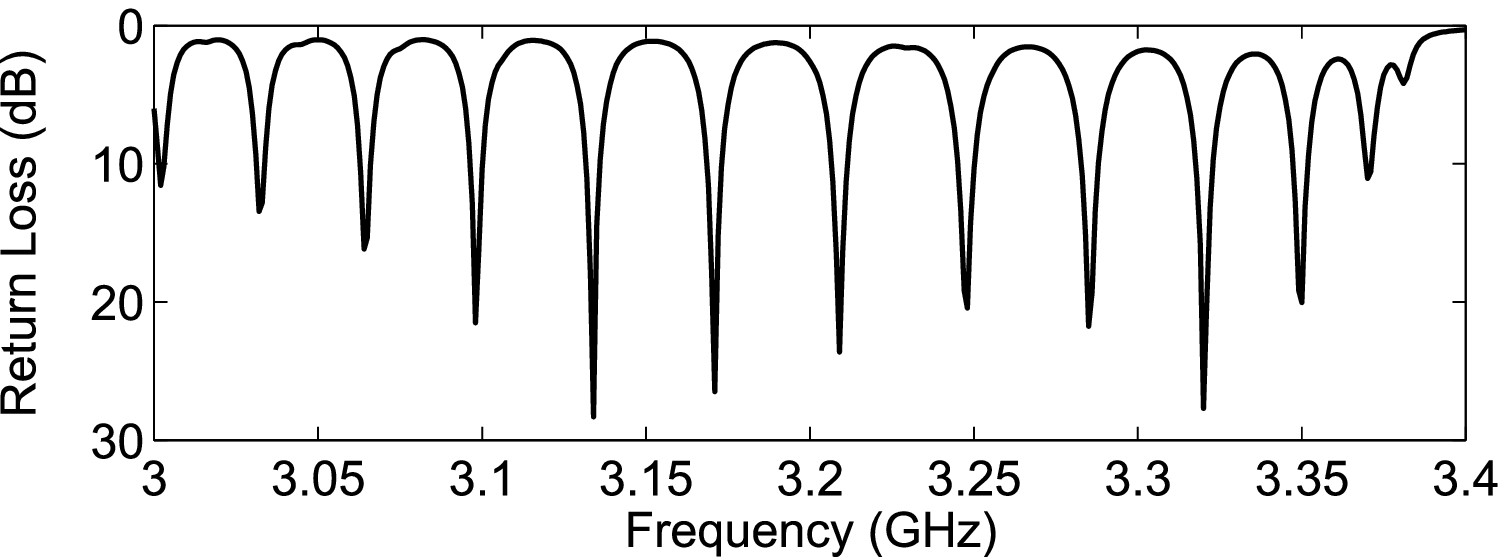}
\label{fig6:subfig2}
}
\\
\subfigure[]{
\includegraphics[width=2.8in]{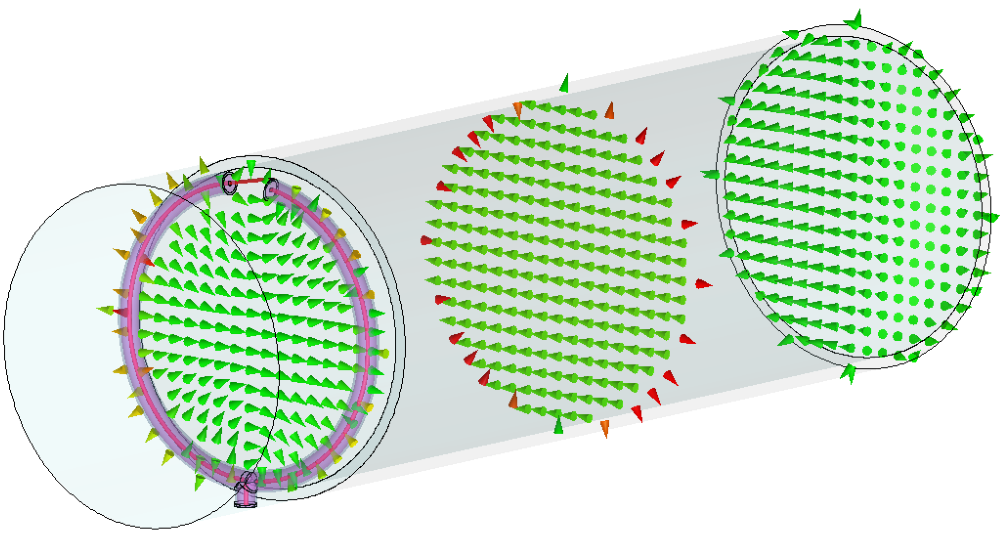}
\label{fig6:subfig5}
}
\subfigure[]{
\includegraphics[width=1.5in]{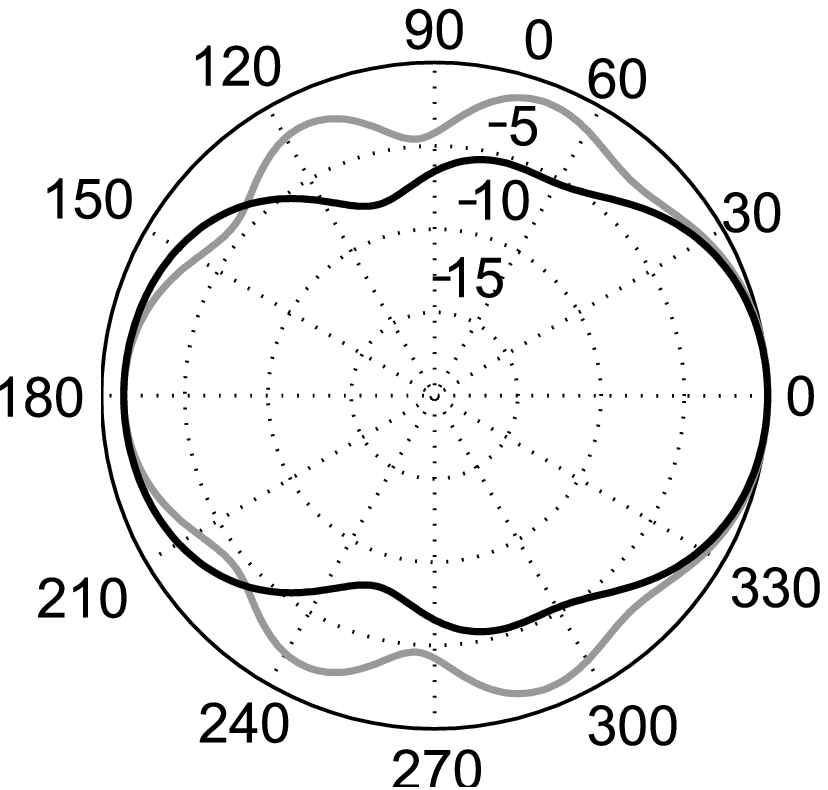}
\label{fig6:subfig6}
}
\caption{\unskip (a) Full-wave simulation model of the shielded-loop-excited OEWG and (b) the resulting return loss. (c) Complex electric-field vectors at different planes in the miniaturized OEWG probe and (d) normalized directivity patterns in the H-plane (black curve) and E-plane (grey curve) at $f=3.320$GHz.}
\vspace{-5pt}
\end{figure}

\subsection{Shielded Loop Excitation}
The standard approach for exciting the {\em TE}$_{11}$ mode in an above-cutoff circular waveguide involves a coaxial pin/loop extending from the side/back wall. In a metamaterial-lined circular waveguide operating below its natural cutoff, the potential for such a source to couple into the full spectrum of frequency-reduced modes must be managed. For instance, preliminary simulations of pin excitations located in or very close to the lined waveguide section revealed that other-order modes such as the {\em EH}$_{01}$ and {\em HE}$_{21}$ are strongly excited below $f_{01}$ and $f_{21}$, respectively, resulting in unexpected locations in the return-loss peaks and undesired far-field patterns.

To overcome these issues, a balanced shielded loop~\cite{whiteside1964loop} is placed within a closed evanescent waveguide section of length $L_1$, as shown in Fig.~\ref{fig6:subfig1}. To maximize compactness, this section is given an outer radius of $b_1=15$mm and is smoothly connected to the metamaterial-lined OEWG section described in Sec.~\ref{sec:theory}. The shielded loop of radius $c_r$ is placed at a distance $p_z$ from a PEC back wall and is composed of a teflon-filled $50\Omega$ coaxial transmission line with inner- and outer-conductor radii of $0.29$mm and $0.93$mm, respectively. The outer conductor and teflon are stripped over an angle $\theta_g=20^{\circ}$, and the inner conductor is shorted to the outer conductor across this gap. The balanced nature of the shield currents on the loop minimizes coupling to the longitudinal ($z$-directed) electric fields. Instead, it creates a longitudinal magnetic field that, in combination with the loop's size, ensures strong coupling to the {\em HE}$_{11}$ mode and weak interaction with the {\em EH}$_{01}$ and {\em HE}$_{21}$ modes. Since loss plays an important role in matching, only the lossy metamaterial case is studied for this arrangement.

After an optimization procedure, it was determined that $L_1=17.8$mm, $p_z=15.0$mm, and $c_r=12.8$mm provided a significantly improved return loss ($>10$dB) at each resonant peak, as shown in Fig.~\ref{fig6:subfig2}, while still maintaining a return loss greater than $1$dB across the band. Consider the resonant peak at $f=3.320$GHz. Figure~\ref{fig6:subfig5} shows the electric field vectors and complex magnitudes at this frequency taken at three locations in the waveguide: at the shielded loop, at the center of the metamaterial-lined region, and at the open aperture. The field patterns confirm that the shielded loop strongly excites the {\em TE}$_{11}$ mode, which is coupled to the {\em HE}$_{11}$ mode in the metamaterial-lined waveguide. Shown in Fig.~\ref{fig6:subfig6} are the H-plane (black curve) and E-plane (grey curve) normalized directivity patterns at $f=3.320$GHz. The far-field features are similar to the waveport-excited `MTM-Lined Lossy' case, but with a reduced directivity of $3.43$dB and a F/B ratio of $1.38$dB. This is expected with the elimination of the larger excitation waveguide, whose front face served as a reflector. Indeed, it has been observed that the directivity of the miniaturized probe may be increased by re-introducing a reflecting plate behind the open aperture. These field and radiation patterns also exhibit a linear polarization purity of greater than $35$dB, which is consistent over the entire band. The antenna parameters for the shielded-loop-excited case are summarized in Table II.

\begin{table}[t]
\begin{center}\label{table:2}
\begin{tabular}{cccccc}
\multicolumn{6}{c}{TABLE II: \textsc{Shielded-Loop-Excited Antenna Parameters}} \\
Direct. (dB)&$\eta_{rad}$&Gain (dB)&HPBW$_{H}$&HPBW$_{E}$&F/B (dB)\\
\hline
3.43&28.8&-1.94&82.2&168.0&1.38\\
\lasthline
\end{tabular}
\end{center}
\vspace{-5pt}
\end{table}

\begin{figure}[!t]
\centering
\includegraphics[width=3.2in]{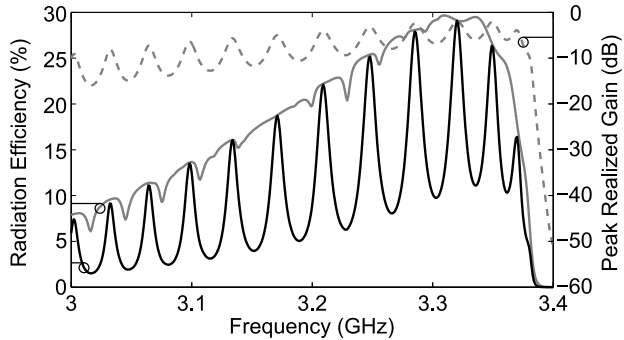}
\caption{\unskip Radiation efficiencies for the shielded-loop-excited OEWG, defined by $\eta_{r,i}=P_{rad}/P_{inc}$ (solid black curve) and $\eta_{r,a}=P_{rad}/P_{acc}$ (solid grey curve). The peak realized gain (dashed grey curve) is also shown.\label{fig8}}
\vspace{-10pt}
\end{figure}

Figure~\ref{fig8} presents the gain and radiation efficiency for the shielded-loop-excited metamaterial-lined OEWG. The following radiation-efficiency definitions are used to aid in discussion: $\eta_{r,i}=P_{rad}/P_{inc}$ (solid black curve) and $\eta_{r,a}=P_{rad}/P_{acc}$ (solid grey curve), in which $P_{rad}$, $P_{acc}$, and $P_{inc}$ are the radiated, accepted, and incident powers, respectively, with respect to the coaxial feed. Whereas $\eta_{r,i}$ captures both return loss and insertion loss due to the liner, $\eta_{r,a}$ captures only insertion loss. Progressing from the {\em HE}$_{11}$ cutoff ($f_{11}=3.381$GHz) down further into the backward-wave band, $\eta_{r,a}$ ramps up to its maximum value of $29.7\%$ at $f=3.310$GHz, after which it experiences a generally linear decrease as frequency reduces. This is consistent with the observed behaviour of $\alpha$ (the modal attenuation due to loss), which also increases towards lower frequencies. However, given the strongly dispersive nature of the metamaterial-lined OEWG, it is clear that $\eta_{r,i}$ is a more realistic definition of radiation efficiency. Indeed, $\eta_{r,i}$ attains local maxima at each of the return-loss peaks shown in Fig.~\ref{fig6:subfig2}, where the matching is most substantially improved.  At $f=3.320$GHz the return loss of $27.7$dB corresponds to a maximum radiation efficiency of $28.8\%$. Note that return loss is maximized at $f=3.134$GHz ($28.3$dB), but the increased insertion loss at these lower frequencies reduces $\eta_{r,i}$ to $16\%$. In general, the insertion loss may be alleviated by choosing a shorter metamaterial-lined waveguide length. However, the waveguide length also establishes the distribution of resonant peaks in the band and, at a minimum, must be large enough to allow any practical periodic implementation of the metamaterial liner to satisfy the definition of an effective medium.
\begin{figure}[!t]
\centering
\includegraphics[width=2.2in]{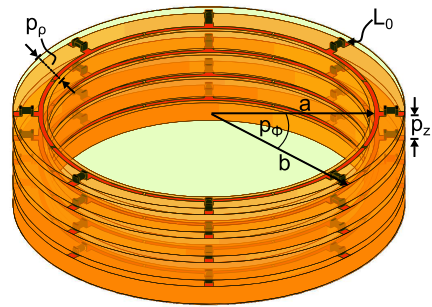}
\caption{\unskip A multilayer printed-circuit implementation of an ENNZ metamaterial liner.\label{fig9}}
\vspace{-5pt}
\end{figure}

The peak realized gain (dashed grey curve) shown in Fig.~\ref{fig8} is defined as the maximum directivity of the antenna at broadside, reduced by $\eta_{r,i}$. A maximum value of $-1.94$dB is achieved at $f=3.320$GHz. This represents a vast improvement over a similarly sized unlined waveguide excited with the same shielded loop, which had a gain of $-76.1$dB and for which virtually no power is radiated. In fact, the gain of the metamaterial-lined OEWG is at least $60$dB greater than that of the unlined waveguide everywhere over the band. Another meaningful comparison may be drawn to a PEC dipole with a length $l$ equal to the diameter of the `MTM-Lined' OEWG (i.e., $l=2b_2=30$mm). The directivity and peak realized gain of a $30$mm dipole fed by a $50\Omega$ source at $f=3.320$GHz are confirmed through full-wave simulations to be $2.24$dB and $-7.69$dB, respectively. Therefore, the miniaturized `MTM-Lined' OEWG offers a nearly four-fold improvement in gain over a similarly sized dipole. Furthermore, it is up to four times smaller than the capacitively loaded evanescent waveguide probe mentioned in Sec.~\ref{sec:intro}. The gain of the metamaterial-lined OEWG may be further improved by sacrificing the degree of miniaturization (over $42\%$ in the studied case), which would result in lower insertion loss due to less field confinement in the liner and higher directivity due to larger aperture size.
\section{Practical Realization of Metamaterial Liners}
\label{sec:realization}
In the previous sections, the liner was modeled as an effective medium with an isotropic and homogeneous permittivity. However, practical metamaterials are often anisotropic. Therefore, it is worth investigating if similar below-cutoff propagation/radiation characteristics are observed for practical anisotropic liners, one variety of which may be constructed using an array of thin wires. Thin-wire media exhibit a Drude-like permittivity dispersion for electric fields aligned parallel to the wire axes and are a cost-effective realization of low-frequency negative permittivities. They also benefit from being easily fabricated using traditional printed-circuit techniques.
While rectangular thin-wire grids have been used to improve the bandwidth and/or gain of rectangular and conical horn antennas~\cite{ramaccia2013,lier2011octave}, the circular waveguide suggests the use of a cylindrical arrangement.

Figure~\ref{fig9} presents a practical, anisotropic thin-wire metamaterial design, which consists of a multilayer arrangement of azimuthally and radially directed copper traces on a dielectric substrate. The metamaterial unit cell has thickness $p_\rho$ in the $\rho$ direction and periods $p_\phi$, and $p_z$ in the $\phi$, and $z$ directions, respectively. When the traces are unloaded they only weakly affect the waveguide environment and do not exhibit the strong inductances needed to achieve ENNZ properties. This can be attributed to their finite lengths in $\rho$ and $\phi$, and the additional boundary conditions imposed at the PEC and vacuum boundaries~\cite{silveirinha2008abc}. In order to restore their negative-permittivity response, the $\rho$-directed traces are loaded discretely with inductors of value $L_0$~\cite{pendry1996}. Capacitive gaps are placed symmetrically between each $\rho$-directed inductor in the $\phi$-directed traces to prevent unwanted resonances in the $\phi$ direction. No $z$-directed wires are employed, so as to avoid the propagation of {\em TEM} modes~\cite{silveirinha2005}.

\begin{figure}[!t]
\centering
\includegraphics[width=3in]{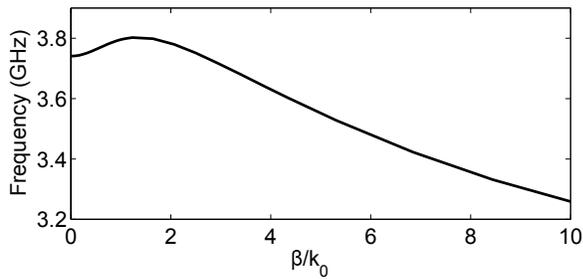}
\caption{\unskip Dispersion of the frequency-reduced {\em HE}$_{11}$ mode for the thin-wire ENNZ liner shown in Fig.~\ref{fig9}.\label{fig10}}
\vspace{-5pt}
\end{figure}
Full-wave eigenmode simulations of the structure shown in Fig.~\ref{fig9} employing representative values of $p_\rho$, $p_\phi$, $p_z$, $L_0$ and all practical sources of dielectric, conductor, and component losses reveal the desired frequency-reduced {\em HE}$_{11}$ mode, whose dispersion is shown in Fig.~\ref{fig10}. This structure is strictly anisotropic, and may be homogenized using an anisotropic effective permittivity tensor in a cylindrical coordinate system~\cite{pollockAEC2013}. Nevertheless, careful design of its geometrical parameters and discrete loading yields dispersion features and miniaturization properties similar to those of an isotropic ENNZ liner.

\section{Conclusion}
\label{sec:conclusion}
This work has presented a study of the radiation characteristics of metamaterial-lined OEWG probe antennas whose aperture dimensions have been miniaturized. It was shown that the introduction of a thin ENNZ liner into the OEWG volume results in a frequency-reduced backward-wave passband well below its natural cutoff frequency. Full-wave simulations of the metamaterial-lined OEWG probe in the frequency-reduced region reveal multiple resonant return-loss peaks where the miniaturized antenna exhibits a substantially improved gain over both a similarly sized (and below-cutoff) circular waveguide probe and a reference dipole. An easily fabricated anisotropic metamaterial-liner implementation based on inductively loaded printed metallic traces is shown to exhibit dispersion and miniaturization properties consistent with those observed for isotropic, homogeneous ENNZ liners.

\ifCLASSOPTIONcaptionsoff
  \newpage
\fi

\bibliography{AP1405-0669-R2-BIBLIOGRAPHY}
\bibliographystyle{IEEEtran}

\end{document}